\documentclass[rmp,nobibnotes,floats]{revtex4}
\usepackage{amsmath}
\usepackage{graphicx}
\bibpunct{(}{)}{;}{a}{,}{,}

\begin{document}

\title{Steady-state mode I cracks in a viscoelastic  triangular lattice}

\author{Leonid Pechenik and  Herbert Levine}
\affiliation{Department of Physics\ \ \ \ \ \ \ \ \ \ \ \ \ \ \ \ \ \ \ \ \ \ \ \ \ \ \ \ \ \ \ \ \ \ \ \\
University of California, San Diego\ \ \ \ \ \ \ \ \ \ \ \ \ \ \ \ \ \ \ \ \ \ \ \ \ \ \ \ \ \ \ \ \ \ \ \ \\ 
La Jolla, CA  92093-0319  USA}
\author{David A. Kessler}
\thanks{Corresponding Author; FAX: +972-3-535-3298}
\email{kessler@dave.ph.biu.ac.il}
\affiliation{Department of Physics\ \ \ \ \ \ \ \ \ \ \ \ \ \ \ \ \ \ \ \ \ \ \ \ \ \ \ \ \ \ \ \ \ \ \ \ \ \ \\
Bar-Ilan University\ \ \ \ \ \ \ \ \ \ \ \ \ \ \ \ \ \ \ \ \ \ \ \ \ \ \ \ \ \ \ \ \ \ \ \ \ \ \ \ \ \ \ \ \ \ \ \\ Ramat-Gan, Israel}

\begin{abstract}
We continue our study of the exact solutions for steady-state cracks
in ideally brittle viscoelastic lattice models by focusing on
mode I in a triangular system. The issues we address
include the crack velocity versus driving curve as well as the onset
of additional bond-breaking, signaling the emergence of complex
spatio-temporal behavior. Somewhat surprisingly, the critical
velocity for this transition becomes a decreasing function of the
dissipation for sufficiently large values thereof. At the end, we
briefly discuss the possible relevance of our findings for experiments
on mode I crack instabilities.\newline\newline
\end{abstract}

\maketitle

\section{Introduction}

In the previous paper \citep{plk-old}, we showed how one could extend the 
methodology
originally devised by \citet{slepyan} and obtain a closed form 
solution for
mode III cracks propagating steadily in a square lattice in
the presence of dissipation in the form of a Kelvin viscosity. This
approach allowed us to determine the effect of dissipation on
the crack velocity as a function of driving. Similarly, we found
the critical
velocity (and the associated critical time) at which other bonds in
the lattice have displacements larger than the assumed breaking
criterion, signaling an instability in the steadily-propagating crack
and the onset of more complex spatio-temporal behavior.

In this second paper, we extend this analysis to the more interesting
(and more experimentally relevant) case of mode I cracks. We now work
on a triangular lattice and assume that there are 
central force ideally brittle viscoelastic springs connecting particles at the
lattice sites. Again, we use ideas borrowed from \citet{slepyan2} to
solve this problem exactly on an infinite lattice. We also solve
the same system numerically on a lattice of finite transverse extent
so as to provide an independent check on some of the results. 

Perhaps the most interesting finding reported here concerns the aforementioned
onset of additional bond-breaking at large enough velocity. At small
dissipation, diagonal bonds that are offset from the assumed crack line in
the vertical direction are the most ``dangerous". The critical
velocity at which one of these bonds exceeds critical displacement is
more or less independent of the dissipation at low values thereof and
eventually rises sharply as the system becomes increasingly damped.
This is similar to the behavior obtained for mode III.
At moderate to large damping, however, horizontal bonds become
more relevant and eventually give rise to a critical velocity which decreases
with increasing damping. At the end, we will comment on the possible
relevance of our results for trying to make sense of recent 
experiments \citep{exp1,exp2,exp3,exp4}
and simulations on instabilities during dynamic fracture.

\section{Mode I on a triangular lattice}

Our model consists of central force springs between mass points
located on a triangular lattice. We introduce unit vectors in the 
lattice directions as follows: $\hat{d}_1$ points in the direction of 
the $x$-axis and each of the subsequent unit vectors $\hat{d}_i$ with
$2 \leq i \leq 6$ rotated by an additional $\pi/3$ in the counter-clockwise 
direction. This means that $\hat{d} _{i+3} = -\hat{d} _{i}$.
We will always assume that the steady-state crack breaks bonds between
the rows with $y=0$ and $y=\sqrt 3 /2$, as it moves in the 
direction of  the $x$-axis. 

As already discussed in our previous paper on mode III cracks, each
bond is taken to be linear until a critical displacement at which
point it completely breaks. Also, dissipation is incorporated as
a Kelvin viscosity term. Given these assumptions, the
equation describing the  motion of the masses in the lattice is
\begin{equation}\label{eqn1}
(1+\eta\frac{\partial}{\partial t})\sum_{i=1}^6 Q_i(t,\vec {x} ) 
\hat{d}_i - 
\frac{\partial^2 \vec{u}(t,\vec{x})}{\partial t^2}= -\vec{\sigma}(t,\vec{x}) \ .
\end{equation}
Here, $\vec{x}=(x, y)$ is a lattice vector in the plane and 
$Q_i(t,\vec{x})$ is the elongation of bond $i$ emanating
from the specific lattice site, in the 
approximation that this elongation is much smaller than the unstretched
bond length,
\begin{equation}
Q_i(t,\vec{x})= \left( \vec{u}(t,\vec{x}+\hat{d}_i)-
\vec{u}(t,\vec{x}) \right) \cdot \hat{d}_i \ .
\end{equation}
Finally, the driving term  $\vec{\sigma}(t, \vec{x})$
consists of forces which compensate for the forces from broken bonds (which
are included on the left-hand side) as well as
any external forces  acting on the crack boundaries.

Next we suppose that the crack is moving with a constant speed $v$. This
allows us to change variables to $\tau=x-v t$, $\vec{z}=(\tau, y)$, which 
gives
\begin{equation}\label{eqv}
(1 - \eta v \frac{\partial}{\partial \tau}) \sum_{i=1}^6 Q_i( \vec{z} ) 
\hat{d}_i -v^2 
\frac{\partial^2 \vec{u}(\vec{z})}{\partial \tau^2}= -\vec{\sigma}(\vec{z}) \ .
\end{equation}
Because of the form of $Q_i(\vec{z})$, we have the obvious symmetries
\begin{equation}
Q_2(\vec{z})=Q_5(\vec{z}+\hat{d}_2), \; Q_3(\vec{z})=Q_6(\vec{z}+\hat{d}_3) \ .
\end{equation}
We will impose one additional symmetry condition on the $Q_i$. As the
crack propagates, it alternately breaks bonds in the $\hat{d}_2$
and $\hat{d}_3$ directions. Let us define our coordinates such  that at 
$\tau=0$ and $t=0$, (so that $x=0$), the 
crack breaks a bond in the direction of $\hat{d}_2$. This means that for 
$\tau>0$ this bond ($Q_2(\vec{z}_0)$ with $\vec{z}_0=(\tau,0)$) is unbroken 
and for $\tau<0$ it is always broken. If
we shift one lattice spacing to the left, the $\hat{d}_2$ bond at $x=-1$
will break at
$t=-1/v$ (here also $\tau = 0$ in $Q_2(\vec{z}_0)$).  It
is clear that the bond in the $\hat{d}_3$ direction at $x=0$ breaks at a time-point
between these two events; we will assume that
this time is exactly at the midpoint,  $t =-1/(2 v) $
or equivalently $\tau = 1/2$
in $Q_3(\vec{z}_0)$. This means that for
$\tau>1/2$ this bond is unbroken and for $\tau<1/2$ it is always broken.
We will assume an even more strict symmetry condition for these bonds, namely 
\begin{equation}\label{cond}
Q_2(\vec{z}_0)=Q_3(x-v(t-\frac{1}{2v}),y=0)= 
Q_3(\vec{z}_0+\frac{1}{2}\hat{d}_1)
\end{equation}
which will need to 
be checked from our solution later.

Let us denote $Q_2(\vec{z}_0) \equiv Q(\tau)$.
Then, we can write down an explicit expression for the forces
$\vec{\sigma}(z)$ on the right hand side of Eq. \ref{eqv}. For the row with $y=0$
\begin{eqnarray}\label{force01}
\vec{\sigma}(\vec{z}_0)&=&\theta(-\tau)\left(\vec{\sigma}_e(\vec{z}_0)-
(1-\eta v \frac{\partial}{\partial \tau}) 
Q_2(\vec{z}_0) \hat{d}_2 \right ) \nonumber \\
&\ &+ \theta(-\tau+\frac{1}{2})\left (\vec{\sigma}_e \, '(\vec{z}_0)-
(1-\eta v \frac{\partial}{\partial \tau}) Q_3(\vec{z}_0)
\hat{d}_3 \right) \\ 
\label{force02}
&=& \theta(-\tau) \left (-P_0(\tau)-
(1-\eta v \frac{\partial}{\partial \tau})Q(\tau)\right )
\hat{d}_2   \nonumber \\
&\ &+\theta(-\tau+\frac{1}{2}) \left ( -P_0(\tau-\frac{1}{2} ) 
-(1-\eta v \frac{\partial}{\partial \tau}) Q(\tau-\frac{1}{2}) \right )
\hat{d}_3 \ . 
\end{eqnarray}
Here $\vec{\sigma}_e$ and $\vec{\sigma}_e \, '$ are the assumed external
forces. In the second form, these have been taken to match the
vectorial nature and also the
symmetry of the bond-breaking terms. The (scalar) $P_0$ will be
specified later. Note that
the  second theta-function in (\ref{force01}) reflects the
already mentioned fact that the bond from point $\vec{z}_0$
in the $\hat{d}_3$ direction breaks earlier (by time interval $1/2v$) than 
the bond from the same
point in the $\hat{d}_2$ direction. 

Let us write down the corresponding force for the row $y=\sqrt{3}/2$.
For the point $\vec{x}=\hat{d}_2$, the 
bond in the $\hat{d}_5$ direction is physically the same as the $\hat{d}_2$
bond at $\vec{x}=(0,0)$ and therefore breaks at $\tau=0$ in 
$Q_5(\vec{z}_0+\hat{d}_2)=Q_2(\vec{z}_0)$.
Now, the $\hat{d}_6$ bond at $\vec{x}=\hat{d_2}$ breaks later,
at $\tau=-1/2$ in $Q_6(\vec{z}_0+\hat{d}_2)$. Thus we have
\begin{eqnarray}\label{force11}
\vec{\sigma}(\vec{z}_0+ \hat{d}_2)&=&\theta(-\tau)\left (\vec{\sigma}_e(\vec{z}_0+\hat{d}_2)-
(1-\eta v \frac{\partial}{\partial \tau}) 
Q_5(\vec{z}_0+\hat{d}_2) \hat{d}_5 \right ) \nonumber \\
&\ &+\theta(-\tau-\frac{1}{2})\left (\vec{\sigma}_e'(\vec{z}_0+\hat{d}_2)-
(1-\eta v \frac{\partial}{\partial \tau}) Q_6(\vec{z}_0+\hat{d}_2)
\hat{d}_6 \right ) \\ 
\label{force12}
&=&\theta(-\tau)\left(-P_0(\tau)-(1-\eta v 
\frac{\partial}{\partial \tau})Q(\tau)\right )
\hat{d}_5  + \nonumber \\
&\ &+\theta(-\tau-\frac{1}{2}) \left ( -P_0(\tau+\frac{1}{2}) 
-(1-\eta v \frac{\partial}{\partial \tau}) Q(\tau+\frac{1}{2}) \right )
\hat{d}_6 \ .
\end{eqnarray}
where we used the symmetry in Eq. \ref{cond} to write $Q_6(\vec{z}_0+\hat{d}_2)=
Q_3(\vec{z}_0+\hat{d}_2-\hat{d}_3)=Q_2(\vec{z}_0+\hat{d}_2-\hat{d}_3-\hat{d}_1/2)=
Q(\tau+1/2)$. 
Finally, we can rewrite all these forces in a more compact form,
\begin{eqnarray}\label{forcen0}
\vec{\sigma}(\vec{z}_0)&=&N(\tau)\hat{d}_5+N(\tau-\frac{1}{2})\hat{d}_6 \ ,\\ 
\label{forcen1}
\vec{\sigma}(\vec{z}_0+\hat{d}_2)&=&N(\tau)\hat{d}_2+N(\tau+\frac{1}{2})\hat{d}_3
\ ,
\end{eqnarray}
where 
\begin{equation}
N(\tau)=\theta(-\tau)\left(P_0(\tau)+(1-\eta v 
\frac{\partial}{\partial \tau})Q(\tau) \right ) \ .
\end{equation}

Next, we proceed by  doing a Fourier transformation of Eq. (\ref{eqv}), over
the continuous variable $\tau$ and the
discrete variable $y=n \sqrt 3/2 $, with integer $n$. In detail,
\begin{subequations}
\begin{eqnarray} \label{Fourdir}
u^F(\tau ,s)&=&\sum_{n=-\infty}^{+\infty}u\left( \tau ,\frac{\sqrt 3}{2}n\right)
e^{i \frac{\sqrt 3}{2} s n} \ ,\\
u(\tau ,y)&=&\frac{\sqrt 3}{4 \pi} \int_{-\frac{2 \pi}{\sqrt 3}}
^{\frac{2 \pi}{\sqrt 3}}u^F(\tau ,s)e^{ - i s y} d s \ ;
\end{eqnarray}
\end{subequations}
\begin{subequations}
\begin{eqnarray} \label{Fourier}
u^{FF}(q,s) &=&\int_{-\infty}^{+\infty}u^F(\tau ,s)e^{i q \tau}d \tau \ , \\
u^F(\tau ,s)&=&\frac{1}{2\pi}\int_{-\infty}^{+\infty}u^{FF}(q,s)e^{-i q \tau}dq
\ .
\end{eqnarray}
\end{subequations}
We thereby obtain
\begin{equation} \label{Foureq}
- v^2 q^2 \vec{u}^{FF}-(1+i \eta v q )\sum_i Q_i^{FF}\hat{d}_{i}
=\vec{\sigma}^{FF}\ ,
\end{equation}
with
\begin{equation}
Q_i^{FF}=(e^{-i \vec{r} \cdot \hat{d}_{i}}-1)\vec{u}^{FF} \cdot \hat{d}_{i}
\end{equation}
and $\vec{r}=(q,s)$. 

To find $\vec{\sigma}^{FF}$ we first transform Eqs. (\ref{forcen0},  \ref{forcen1})
over $\tau$
\begin{subequations}
\begin{eqnarray}
\vec{\sigma}^F(q,n=0)&=&N(q)\hat{d}_{5} + e^{i \frac{q}{2}} N(q) \hat{d}_{6} \ ,\\
\vec{\sigma}^F(q,n=1) e^{- i \frac{q}{2}}&=&N(q) \hat{d}_{2} + 
e^{- i \frac{q}{2}} N(q) \hat{d}_{3} \ ,
\end{eqnarray}
\end{subequations}
with
\begin{equation}\label{nq}
N(q)=P_0^- + (1+i \eta v q) Q^- - \eta v Q(0) \ .
\end{equation}
The superscript ``$-$'' refers to the part of the Fourier transform
which is analytic in the lower half plane of the variable $q$; our notation 
follows that in our previous mode III paper.
Finally, using (\ref{Fourdir}) we get
\begin{equation}
\vec{\sigma}^{FF}=N(q)\hat{d}_{5} + e^{i \frac{q}{2}} N(q) \hat{d}_{6}+
\left ( e^{ i \frac{q}{2}} N(q) \hat{d}_{2} + N(q) \hat{d}_{3} \right) 
e^{ i s \frac{\sqrt 3}{2}} \ .
\end{equation}
This then completes the derivation of the basic equation that needs
to be solved to determine the steady-state crack field.

\section{Wiener-Hopf Solution}

We wish to use the Wiener-Hopf technique to solve the equation derived
above for the elongation field $Q$. To start, we note that
Eq.  (\ref{Foureq}) can be written in matrix form 
\begin{equation}\label{matrixeq}
{\bf M}  \vec{u}^{FF} = \vec{\sigma}^{FF}
\end{equation} 
with the 2x2 matrix
\begin{equation}
\bf{M} =\left(
\begin{array}{clrr}
A_1 & A_3 \\
A_3 & A_2
\end{array} \right)
\end{equation}
where
\begin{subequations}
\begin{eqnarray}
A_1&=&-v^2 q^2 + (1+ i \eta v q) \left(4 \sin^2 \frac{\vec{r} \cdot \hat{d}_{1}}{2}+
 \sin^2 \frac{\vec{r} \cdot \hat{d}_{2}}{2} +  \sin^2 \frac{\vec{r} \cdot \hat{d}_{3}}{2} \right) \ , \\
A_2&=&-v^2 q^2 + 3 (1+ i \eta v q) \left(\sin^2 \frac{\vec{r} \cdot \hat{d}_{2}}{2} +  
\sin^2 \frac{\vec{r} \cdot \hat{d}_{3}}{2}\right) \ , \\
A_3&=&\sqrt 3 (1+ i \eta v q) \left(\sin^2 \frac{\vec{r} \cdot \hat{d}_{2}}{2} -
\sin^2 \frac{\vec{r} \cdot \hat{d}_{3}}{2} \right)  \ ,
\end{eqnarray}
\end{subequations}
and
\begin{equation}
\vec{\sigma}^{FF}=-\frac{N(q)}{2} \left(
\begin{array}{clrr} 
(1- e^{i \frac{q}{2}})(1 + e^{ i \frac{\sqrt 3}{2}s} ) \\
\sqrt 3 (1+ e^{i \frac{q}{2}})(1 - e^{ i \frac{\sqrt 3}{2}s }) 
\end{array} \right) \ .
\end{equation}
The solution of (\ref{matrixeq}) is therefore
\begin{equation}\label{solution}
\vec{u}^{FF}=\frac{1}{\det {\bf M}}\left(
\begin{array}{clrr}
\det {\bf M}_1 \\
\det {\bf M}_2
\end{array} \right)
\end{equation}
where
\begin{equation}
{\bf M}_1=\left(
\begin{array}{clrr}
\sigma^{FF}_x & \sigma^{FF}_y \\
A_3 & A_2
\end{array} \right) \ ,\; \; \; \; 
{\bf M}_2=\left(
\begin{array}{clrr}
A_1 & A_3 \\
\sigma^{FF}_x & \sigma^{FF}_y 
\end{array} \right) \ .
\end{equation}

Now, we want to extract from the solution (\ref{solution}) an equation for the
variable we are interested in, namely $Q^F(q)$. From the definition of $Q$
we have
\begin{equation}\label{qf1}
Q^F(q)=Q_2^F(q, n=0)=\frac{\sqrt 3}{4 \pi} \int_{-\frac{2 \pi}{\sqrt 3}}
^{\frac{2 \pi}{\sqrt 3}} (e^{-i \vec{r} \cdot \hat{d}_2}-1) 
\vec{u}^{FF} \cdot \hat{d}_2 \ ds \ .
\end{equation}
From the basic solution above, we can evaluate $\vec{u}^{FF} \cdot \hat{d}_2$.
\begin{equation}\label{ud2}
\vec{u}^{FF}  \cdot 
\hat{d}_2=\frac{
\left |
\begin{array}{clrr}
-d_{2x}A_3+d_{2y}A_1 & -d_{2x}A_2+d_{2y}A_3 \\
\sigma^{FF}_x & \sigma^{FF}_y
\end{array} 
\right| } { | {\bf M} |} \ .
\end{equation}

Before proceeding to evaluate the integral, it is worthwhile to pause
and check the basic symmetry conditions. We  find that
\begin{equation}\label{q3}
Q_3^F(q, n=0)=\frac{\sqrt 3}{4 \pi} \int_{-\frac{2 \pi}{\sqrt 3}}
^{\frac{2 \pi}{\sqrt 3}} (e^{-i \vec{r} \cdot \hat{d}_3}-1) 
\vec{u}^{FF} \cdot \hat{d}_3  \  d s \ ,
\end{equation}
where the new dot product is given by
\begin{equation}\label{ud3}
\vec{u}^{FF} \cdot \hat{d}_3=\frac{
\left |
\begin{array}{clrr}
d_{2x}A_3+d_{2y}A_1 & d_{2x}A_2+d_{2y}A_3 \\
\sigma^{FF}_x & \sigma^{FF}_y
\end{array} 
\right| } { | {\bf M} |} \ .
\end{equation}
Now, we can change variables of integration in (\ref{q3}) from $s$ to $-s$.
Under this transformation $A_1|_{s\rightarrow -s}=A_1$, 
$A_2|_{s\rightarrow -s}=A_2$, $A_3|_{s\rightarrow -s}=-A_3$, 
$\sigma_x^{FF}|_{s\rightarrow -s}=e^{-i \sqrt{3} s/2} \sigma_x^{FF}$,
$\sigma_y^{FF}|_{s\rightarrow -s}=-e^{-i \sqrt{3} s/2} \sigma_y^{FF}$,
which gives that $\det {\bf M} |_{s\rightarrow -s}=\det {\bf M}$ and 
$\vec{u}^{FF} \cdot \hat{d}_3 |_{s\rightarrow -s}=-e^{-i \sqrt{3} s/2}
\vec{u}^{FF} \cdot \hat{d}_2$. Therefore (\ref{q3}) changes into
\begin{equation}
Q_3^F(q, n=0)=-\frac{\sqrt 3}{4 \pi} \int_{-\frac{2 \pi}{\sqrt 3}}
^{\frac{2 \pi}{\sqrt 3}} (e^{i q d_{2x}+i s d_{2y}}-1) 
 e^{-i d_{2y} s} \vec{u}^{FF}\hat{d}_2  d s=e^{i \frac{q}{2}}
Q_2^F(q, n=0) \ .
\end{equation} 
This is exactly the  symmetry condition (\ref{cond}); hence, our solution is 
consistent with the assumed symmetry.
We also note here that a similar derivation gives
\begin{equation}\label{q3toq2}
Q_2(\tau, y)=Q_3(\tau+\frac{1}{2},-y)=Q_6(\tau+\frac{1}{2},-y+
\frac{\sqrt 3}{2}) \ .
\end{equation}
More generally, we can 
see what these symmetries mean for  the displacement field $\vec{u}(z)$. 
Using $\det {\bf M}_1 |_{s\rightarrow -s}=e^{-i \sqrt{3} s/2} \det {\bf M}_1$,
$\det {\bf M}_2 |_{s\rightarrow -s}=-e^{-i \sqrt{3} s/2} \det {\bf M}_2$,  we obtain
$u_x^{FF}|_{s\rightarrow -s}=e^{-i \sqrt{3} s/2} u_x^{FF}$ and 
$u_y^{FF}|_{s\rightarrow -s}=-e^{-i \sqrt{3} s/2} u_y^{FF}$. These give
$u_x(\tau,y)=u_x(\tau,-y+\sqrt 3 /2)$ and  
$u_y(\tau,y)=-u_y(\tau,-y+\sqrt 3 /2)$.

We now proceed to calculate the integral in (\ref{qf1}). The 
explicit form of the integrand in (\ref{qf1}) is
\begin{equation}\label{integrand}
 \frac{N(q)}{|{\bf M}|}  \, 
\left| 
\begin{array}{clrr}
3\left(\frac{v^2 q^2}{2}-(1+i \eta v q)(2 \sin^2 \frac{q}{2} +\sin^2 
\frac{ \vec{r} \cdot \hat{d}_3}{2})\right)  & -\frac{v^2 q^2}{2}
+3(1+i \eta v q)\sin^2 \frac{\vec{r} \cdot \hat{d}_3}{2} \\
(\cos \vec{r} \cdot \hat{d}_2-1)-(\cos\frac{\sqrt{3}}{2}s-\cos\frac{q}{2})&
(\cos \vec{r} \cdot \hat{d}_2-1)+(\cos\frac{\sqrt{3}}{2}s-\cos\frac{q}{2})
\end{array}  
\right| \ .
\end{equation}
As $|{\bf M}|$ is an even function of $s$, only the even part of 
the other determinant  contributes to the integral. Then, by the change of 
variable to $w=e^{-i \sqrt 3 s/2}$ we can rewrite (\ref{qf1}) 
as an integral over the unit circle in the complex $w$  plane. After some
tedious algebra, we find that we can write
\begin{equation}\label{denom}
\det {\bf M} = 3 \xi^2 + \alpha \xi + \beta
\end{equation}
where 
\begin{subequations}
\begin{eqnarray} 
\xi &=&(1+i \eta v q) \frac{1}{2} \left(\frac{1}{w}+ w\right) \label{xidef} \ ,\\
\alpha &=& - 2 \cos \frac{q}{2}\left(- 2 v^2 q^2 +3 (1+i \eta v q)(1+ 
2 \sin^2 \frac{q}{2})\right) \ ,\\
\beta &=& 3 (1+i \eta v q)^2 (1 + 3 \sin^2 \frac{q}{2})-4(1+i \eta v q)v^2 q^2
(1 + \sin^2 \frac{q}{2}) + v^4 q^4 \ .
\end{eqnarray} 
\end{subequations}
Similarly, the even part of the determinant in the numerator becomes
\begin{equation}\label{even}
\frac{F(\xi )}{1+i \eta v q} \ ,
\end{equation}
with
\begin{eqnarray}
F(\xi ) &=& 3\left[(\xi -(1+i q \eta v) \cos \frac{q}{2})^2+
2(1+i q \eta v) \sin^2 \frac{q}{2}
( 1 + \cos \frac{q}{2})(1 + i q \eta v - \xi )\right]  \nonumber \\
&\ & -v^2 q^2 \left[(1 + i q \eta v - \xi )( 1 + \cos \frac{q}{2})- \xi \cos \frac{q}{2}
+1 + i \eta v q \right ] \ .
\end{eqnarray}
We write down here also the odd part of this determinant for future reference,
\begin{equation}
G(w)=i ( \frac{1}{w}-w) \sin \frac{q}{2}( v^2 q^2 - 3 (1 + i q \eta v)
\sin^2 \frac{q}{2}) \ .
\end{equation} 

Inserting (\ref{denom}) and (\ref{even}) into 
(\ref{integrand}) we can see that  (\ref{qf1}) becomes
\begin{equation}\label{cplane}
Q^F(q)=\frac{N(q)}{2 \pi i (1 + i \eta v q)} \oint\limits_{|w|=1} 
\frac{F(\xi )}{3 \xi^2 + \alpha \xi + \beta}\frac{dw}{w}
\end{equation}
with integration around the unit circle in the counter-clockwise direction.
The integrand in Eq. (\ref{cplane}) has three poles inside the unit circle; 
i.e., $w=0$ and one $w$
for each of the two roots of the expression (\ref{denom}), 
\begin{equation}
\xi_{1,2}=C\pm\sqrt{C^2-D} \ ,
\end{equation}
with
\begin{subequations}
\begin{eqnarray}
C&=&\cos \frac{q}{2}\left(- \frac{2}{3} v^2 q^2 + (1+i \eta v q)(1+ 
2 \sin^2 \frac{q}{2})\right) \ ,\\
D&=&(1+i \eta v q)^2 (1 + 3 \sin^2 \frac{q}{2})-\frac{4}{3}(1+i \eta v q)v^2 q^2
(1 + \sin^2 \frac{q}{2}) + \frac{1}{3} v^4 q^4 \ .
\end{eqnarray}
\end{subequations}
Then calculating the residues at these poles we get
\begin{equation}\label{qfn}
Q^F(q)=\frac{N(q)}{1+i \eta v q}(1 - S)
\end{equation} 
where 
\begin{equation}\label{Striang}
S(q)=\frac{F(\xi _1)\sqrt{\xi _2^2-(1+i \eta v q)^2}
-F(y_2)\sqrt{\xi _1^2-(1+i \eta v q)^2}}{3(\xi _1-\xi _2)\sqrt{\xi _1^2-(1+i \eta v q)^2}
\sqrt{\xi _2^2-(1+i \eta v q)^2}} \ .
\end{equation}
Note that we are supposed to take those branches of the square roots in 
(\ref{Striang}) which ensure that 
\begin{equation}\label{z12<1}
|w_{1,2}|<1
\end{equation}
with $w_{1,2}$ defined by
\begin{equation}\label{z12}
w_{1,2}=\frac{\xi _{1,2}-\sqrt{\xi _{1,2}^2
-(1+i \eta v q)^2}}{1+i \eta v q}
\end{equation}
If we now substitute $N(q)$ in Eq. (\ref{qfn}) into Eq. (\ref{nq}), we get
exactly the same equation as was obtained for the mode III case in
the first paper \citep{plk-old}
\begin{equation}
Q^+ +S Q^-+\frac{\eta v}{1+ i \eta v q}(1-S)Q(0)=\frac{(1-S)P_0^-}{1+ i \eta v q}
\end{equation}
with $Q$ having the same meaning as $\vee$ in \citep{plk-old}, namely the
elongation of bonds between rows where the crack propagates. 

Therefore, we do not need to repeat any of the discussion from our previous
paper regarding the choice of $P_0^-$ and the factorization into pieces
analytic in the upper and lower half planes. 
Instead, we can just write down the
final answer. For the relationship between the velocity and the driving
displacement, we have
\begin{equation}\label{numres3}
\frac{\Delta}{\Delta_{G}}=\sqrt{\frac{1+\phi \eta v}{A  \phi}}
\exp {\frac{1}{4 \pi i} \int_{-\infty}^{+\infty}\frac{\ln K(\chi)}
{\chi(1+ i \eta v \chi) } d \chi}
\end{equation}  where for this mode I case
\begin{equation}\label{atria}
A=\frac{1}{\sqrt{2} v^2}\left(\frac{\sqrt{\frac{3}{8}-v^2}}{2}
 - \frac{(-\frac{3}{4}+v^2)^2}{\sqrt{3}\sqrt{\frac{9}{8}-v^2}} \right)
\end{equation}
and
\begin{equation}
K(q)=\frac{S^2(q) ( q^2 +\phi^2)}{ A^2 q^2 \phi^2} \ .
\end{equation}
This expression for $A$ contains the transverse speed $c_t=\sqrt{3/8}$ and 
longitudinal speed  $c_l=\sqrt{9/8}$. The zero of $A$ gives the limiting 
speed of crack propagation $c_R=\frac{\sqrt{3-\sqrt3}}{2}=0.563$, which is 
as expected equal to the Raleigh wave speed. 

\begin{figure}\centerline{\includegraphics[width=3.25in]{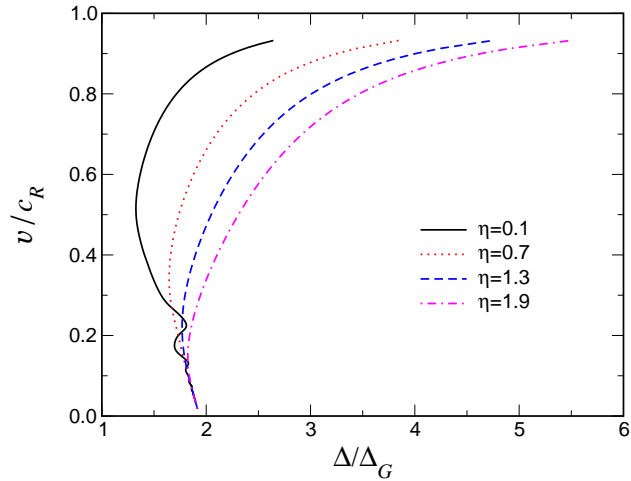}}
\caption{Dependence of 
$\Delta/\Delta_{G}$ on speed $v$ 
for $\eta=0.1$, $0.7$,  $1.3$,  $1.9$ }
\label{figure3}
\end{figure}

\begin{figure}\centerline{\includegraphics[width=3.25in]{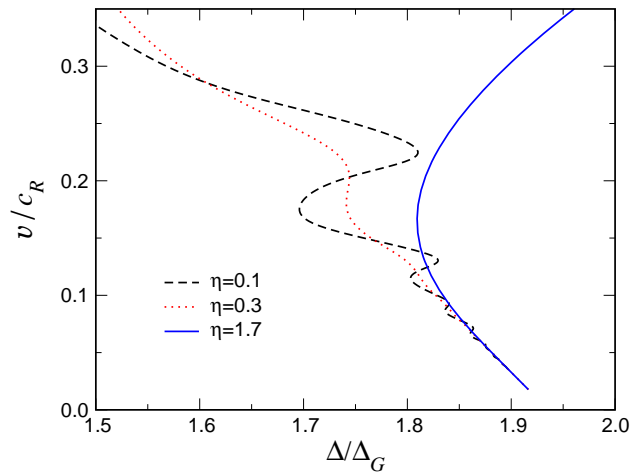}}
\caption{Change of behavior for small $\eta$. Note the rapid oscillations at smaller speeds.}
\label{figure4}
\end{figure}

In our previous paper, we discussed the details of how to
evaluate this equation so as to obtain numerical results for
the crack velocity response curve. These methods can be applied
to this case as well and hence we do not repeat any of the details here.
The results of these calculations are presented in Figs. 
\ref{figure3} and \ref{figure4}. In fact, these findings are quite
similar to the analogous mode III results. Later, we will show that
the upper limiting value of $\Delta/\Delta_G$ for an arrested 
(i.e. non-moving) crack is around $1.94$.
As was the case for mode III, the $\Delta-v$ curves in Figures
\ref{figure3},  \ref{figure4} approach this value as $v$ goes to 0 at any 
value of the damping $\eta$. Slepyan's original calculations, 
done for the dissipationless limit $\eta=0$, show the same asymptotic value.
 
\section{consistency of the steady-state solution}
In this section we calculate the bond displacements in the
vicinity of the crack. This is crucial, as we need to check whether
all bonds assumed to be unbroken in the derivation in fact
have elongations less than unity, the value at which bonds break. 

Let us start with bonds along the crack path, where the elongation
is just determined by transforming $Q^F (q)$ back to physical space. 
This part follows exactly the analogous mode III calculation, and we
easily obtain for $Q^{\pm}$
\begin{subequations}
\begin{eqnarray}
Q^+&=&\frac{Q (0)}{(-iq+0)(1 + i q \eta v)}\frac{S^+}{\left.S^+ 
\right| _{q=\frac{i}{\eta v}}}-\frac{ \eta v Q(0)}{1 + i q \eta v} \ , \\
Q^-&=& \frac{Q (0)}{\left.S^+ 
\right| _{q=\frac{i}{\eta v}}S^- }\frac{1}{(iq+0)(1 + i q \eta v)} 
+\frac{ \eta v Q (0)}{1 + i q \eta v} \ ,
\end{eqnarray}
\end{subequations}
and for $Q(\tau)$
\begin{subequations}
\begin{eqnarray}\label{qpos}
Q(\tau)&=&Q(0)\int_{-\infty}^{+\infty}\frac{ \sqrt{ A \phi} 
e^{- i q \tau}  }
{\sqrt{(i q - 0)( i q - \phi)}  ( 1 + i q \eta v )}\frac{(K^+)^
{\frac{1}{2}} }{\left.S^+ 
\right| _{q=\frac{i}{\eta v}}  }\; \frac{d \, q}{2 \pi}, \hbox{
\ \ \    for $\tau>0$} \\
\label{qneg}
Q(\tau)&=&
Q(0) \int_{-\infty}^{+\infty}\frac{ e^{- i q \tau}}
{\sqrt{A \phi} (i q + 0 )( 1 + i q \eta v )}\left(\frac{i q +\phi}{
iq+0}\right)^{\frac{1}{2}} \frac{(K^-)^
{-\frac{1}{2}} }{\left.S^+ 
\right| _{q=\frac{i}{\eta v}}  }\; \frac{d \, q}{2 \pi} \nonumber \\
&\ &+Q(0) e^{\frac{\tau}{\eta v}}
, \hbox{
\ \ \ \ \ \ \    for $\tau<0$} \ .
\end{eqnarray}
\end{subequations}

\begin{figure}\centerline{\includegraphics[width=3.25in]{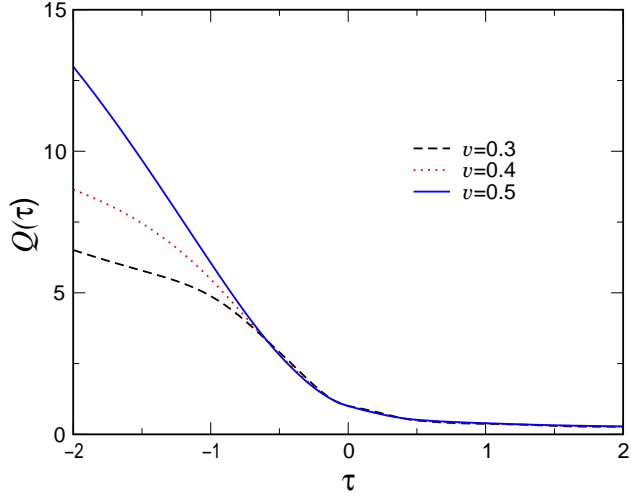}}
\caption{Elongation of bonds on the crack surface, $Q(\tau)$
for $\eta=0.5$.}
\label{figure9}
\end{figure}

The numerical evaluation of these expressions follows the
same methodology as described for mode III.
Typical results are shown in Fig. \ref{figure9}.  We note that the
elongation along the crack line is rather similar to the
same object in the  mode III case. In that case, however, this function
also determined the horizontal bond elongations by simple subtraction.
This is no longer true for mode I, since the vectorial nature of
the problem requires that we take a different component of the displacement
(different than the ones which goes into $Q$) to evaluate this
elongation. In detail, we now have for 
\begin{eqnarray}
Q_h(q) \equiv Q_1^F (q, n=0 )=\frac{\sqrt 3}{4 \pi} \int_{-\frac{2 \pi}
{\sqrt 3}}
^{\frac{2 \pi}{\sqrt 3}} (e^{-i r \hat{d}_1}-1) 
\vec{u}^{FF}\hat{d}_1  d s =  \\  \label{qh1}
(e^{-i q}-1)\frac{\sqrt 3}{4 \pi} 
\int_{-\frac{2 \pi} {\sqrt 3}}^{\frac{2 \pi}{\sqrt 3}}
\frac{\det {\bf M}_1}{\det {\bf M}} d s
\end{eqnarray}
where these matrices were defined in the last section.

We now proceed as before to change variables to $w$ and re-write this
expression in terms of the auxiliary variable $\xi$ defined in Eq. 
(\ref{xidef}).
\begin{equation}
\det {\bf M}_1=-\frac {N(q) H (\xi )}{1+ i \eta v q}+ \hbox{term odd in $s$} 
\end{equation}
where
\begin{eqnarray}
H(\xi )&=&\frac{3}{2}(1-e^{i \frac{q}{2}})(1+ i \eta v q  + \xi )(1+ i \eta v q - 
\xi  \cos \frac{q}{2}-\frac{v^2 q^2}{3})  \nonumber \\
&\ & -\frac{3i}{2} \sin \frac{q}{2} (1+e^{i \frac{q}{2}}) ( \xi ^2-(1+ i \eta v q)^2)
\end{eqnarray}
As before, the integrand in  (\ref{qh1}) has three poles. 
Dropping the irrelevant odd term and performing the integral leads
to the expression
\begin{equation}
Q_h(q)= \frac{N(q)}{1 + i \eta v q} (1-e^{- i q} ) \psi_h(q)
\end{equation}
with
\begin{equation}
\psi_h(q)=\frac{ 1- e^{i \frac{q}{2}}}{2} - 
\frac{H(\xi _1) \sqrt{\xi_2^2-(1+ i \eta v q )^2}-
H(\xi_2) \sqrt{\xi_1^2-(1+ i \eta v q )^2}}
{3 (\xi_1-\xi_2) \sqrt{\xi_1^2-(1+ i \eta v q )^2} 
\sqrt{\xi_2^2-(1+ i \eta v q )^2}} \ .
\end{equation}
Branches of the square root satisfy the same 
conditions (\ref{z12<1}), (\ref{z12}), as earlier. 
The function $N(q)$ can be found from Eq. (\ref{qfn})
\begin{equation}\label{nqnew}
N(q)=\frac{(1+i \eta v q) Q^F(q)}{1 - S} \ .
\end{equation}
Our final answer is 
\begin{equation}
Q_h(q)=\frac{Q(0)}{\left. S^+ 
\right | _{q=\frac{i}{\eta v}} S^-}\frac{(1-e^{- i q} )\psi_h(q)}
{(iq+0)(1+i q \eta v)} \ . 
\end{equation}

\begin{figure}\centerline{\includegraphics[width=3.25in]{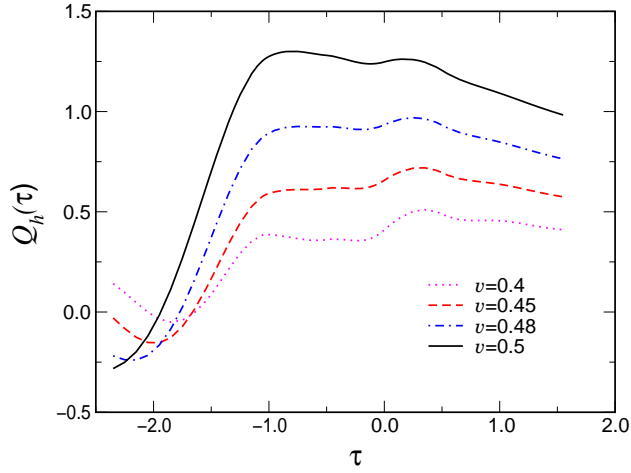}}
\caption{Elongation of the horizontal bonds, $Q_h(\tau)$, for $\eta=0.1$}
\label{figure11}
\end{figure}

For $q$ close to $0$, the  function  $(1-e^{- i q} )\psi_h(q)$ behaves as $q$,
giving a  divergence $1/q^{1/2}$ for  $Q_h(q)$. This is similar
to the divergence in $Q^+(q)$.
Thus, the numerical calculation of $Q_h(\tau)$ is similar to the calculation
 of  $Q(\tau)$ for $\tau>0$. 
Figure \ref{figure11} displays $Q_h(\tau)$ for several values of 
$\eta$ and $v$. Generally speaking, the function $Q_h(\tau)$ has maxima
in two different places, one somewhere in the vicinity of $-1$ and a second
for $\tau>0$. 

\begin{figure}\centerline{\includegraphics[width=3.25in]{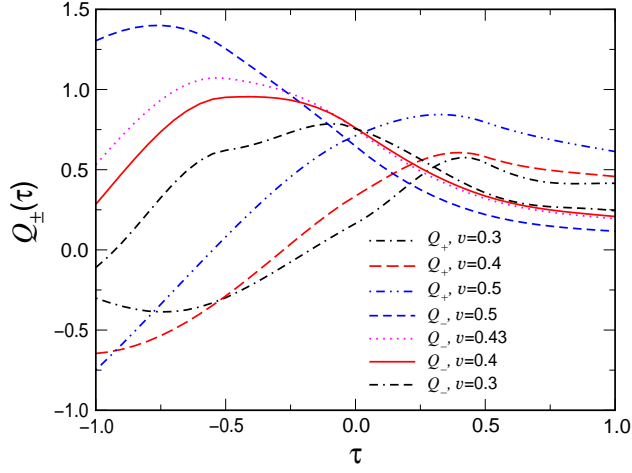}}
\caption{Elongation of bonds on the layer next to the crack layer, 
$Q_\pm(\tau)$ for $\eta=0.1$}
\label{figure10}
\end{figure}

We also need to find the bond elongation between the layers with $n=1$ and 
$n=2$ and
between the layers with $n=-1$ and $n=0$. Due to the symmetry (\ref{q3toq2})
it is sufficient to consider only $Q^F_2(q,n=\pm 1)$, which we 
will denote as $Q_\pm(q)$. We can derive
for it an expression similar to that in Eq. (\ref{cplane}) with the
major difference being that in this case the
odd parts of the explicit  determinant  in Eq. (\ref{integrand}) do not 
cancel out because of the additional factor $w^{\pm 1}$.
\begin{eqnarray} 
Q_\pm(q)&=&\frac{N(q)}{2 \pi i } \oint\limits_{|w|=1} w^{\pm 1}
\frac{\frac{F(\xi )}{1+i \eta v q}+G(w)}{3 \xi ^2 + \alpha \xi + \beta}
\frac{dw}{w} \nonumber \\ \label{q2int}
&=&\frac{N(q)}{2 \pi i } \oint\limits_{|w|=1} 
\frac{\frac{F( \xi )}{1+i \eta v q} \pm G(w)}{3 \xi^2 + \alpha \xi + \beta}
\, dw \ .
\end{eqnarray}
To derive the second equality,  we used a change of variables $w \rightarrow 1/w$,
along with the symmetries of $F$ and $G$. Again the integration in (\ref{q2int})
is in the counter-clockwise direction.
In this case, the integrand in the second form of the integral has just two 
poles inside the unit  sphere, at $w_{1,2}$.
After the by now familiar tedious calculations, we find 
\begin{equation}
Q_\pm(q) = \frac{N(q)}{1+i \eta v q}\psi_\pm (q)
\end{equation}
where
\begin{eqnarray}
\psi_\pm (q)=
-\frac{F(\xi_1) w_1}{3 (\xi_1-\xi_2) \sqrt{\xi_1^2-(1+i \eta v q)^2}} + 
\frac{F(\xi_2) w_2}{3 (\xi_1-\xi_2) \sqrt{\xi_2^2-(1+i \eta v q)^2}}
\nonumber \\
\pm \frac{ g(q) (w_1-w_2)}{3 (\xi_1-\xi_2)}
\end{eqnarray}
with
\begin{equation}
g(q)=- 2 i \sin \frac{q}{2}( v^2 q^2 - 3 (1 + i q \eta v)
\sin^2 \frac{q}{2})
\end{equation}
and $N(q)$ is again given by (\ref{nqnew}).
Finally we obtain
\begin{equation}
Q_\pm(q)=\frac{Q(0)}{\left. S^+ 
\right |_{q=\frac{i}{\eta v}} S^-}\frac{\psi_\pm(q)}{(iq+0)(1+i q \eta v)}
\end{equation}
Just as was the case for $Q_h(q)$ and $Q^+(q)$, 
$Q_\pm(q)$ behaves as $1/q^{1/2}$ near $0$. Thus 
numerical calculations can be performed just as in the previous cases. 
Figure \ref{figure10} 
shows several curves $Q_\pm(\tau)$ for differing parameters. 

\begin{figure}\centerline{\includegraphics[width=3.25in]{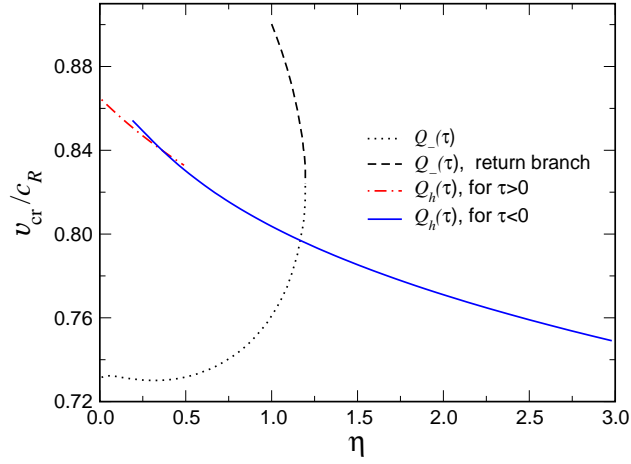}}
\caption{Critical speed $v_{\text cr}$}
\label{figure12}
\end{figure}

\begin{figure}\centerline{\includegraphics[width=3.25in]{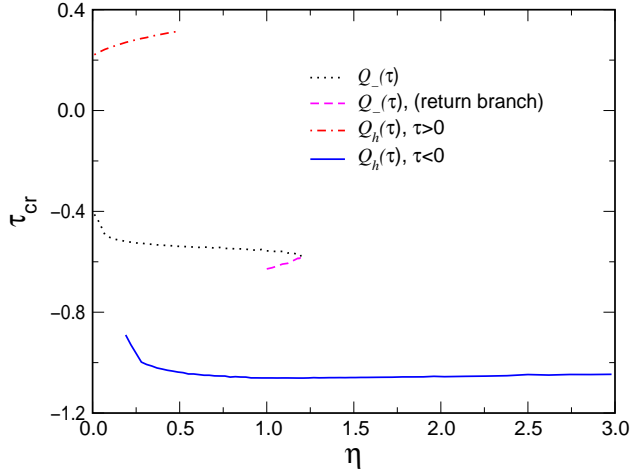}}
\caption{Time of bond breaking at critical speed}
\label{figure13}
\end{figure}

Given the elongations of these ``vulnerable" bonds, we can investigate
the critical speed at which one of these bonds should be broken. 
Figure \ref{figure12} shows the results of our calculations of
this critical speed. In Figure \ref{figure13} we plotted $\tau_{cr}$
at which these bonds break; this will allow us to identify spatial and 
time coordinates of braking and make contact with our finite 
lattice calculations later. 
We found that the maximum value of $Q_-(\tau)$ always reaches 1 before 
$Q_+(\tau)$ and that this is the most dangerous bond for small dissipation.
This curve turns around at an $\eta$ of around $1.2$, so that $Q_-(\tau)$
is always subcritical for larger $\eta$.  This is a result of the
fact that the maximum value of $Q_-(\tau)$ is, surprisingly enough,
not monotonic with velocity, but reaches a maximum and then decreases with
increasing velocity.  The maximum extension of this type of bond occurs,
for large velocity, at some distance from the crack surface.  For small $\eta$,
where the maximum $Q_-(\tau)$ does exceed critical extension, the decrease
of the maximum $Q_-$ with $v$ restabilizes the bond beyond some velocity.
The dominant threshold for $\eta$ above about $1.1$, then,
comes from the horizontal bond breaking. Note that as
$\eta$ increases, the critical velocity decreases. Also, for
small $\eta$, a crossover occurs as the relative importance of the
two different maxima in the horizontal bond elongation reverse; this
region is anyway irrelevant as the next-layer vertical bond breaks at a
lower velocity. Whenever the horizontal bond dominates, the maximum
at  $\tau \simeq -1$ is the governing one. As we see from Figure \ref{figure13}
even for large $\eta$ this critical $\tau$ stays near $-1$, what means
that horizontal bond always breaks near the tip of the crack, 
contrary to the Mode III case, where the point of breaking drifts backward 
with increasing $\eta$.

Many of these features are special to the Mode I problem, and have
no analog in the Mode III calculation.  For Mode III, the horizontal
bond breaking always dominates.  Furthermore, the maximum off-crack surface
bond extension is a strictly increasing function of the velocity.  This
points to the possibility that the dynamics of Mode I may be much richer
than the Mode III dynamics.

\section{Finite lattice model.}
For our mode III calculations, it proved interesting to compare the
exact results derived for an infinite lattice with the numerical
determination of crack propagation properties in lattices of small
transverse size \citep{kl1,dk}. We now discuss similar calculations for our mode I model.
In addition to providing details regarding the size needed to attain
answers relevant to the macroscopic limit (a question of direct relevance for
direct molecular dynamics simulations \citep{md0,md1,md1a,md2,md2a,md3}, for example), 
finite lattice results
can be used as a rough check on some of the predictions obtained above.
Given the complexity of the analysis, having such a check is quite
useful.

\subsection{Arrested crack}

Let us start with trying to find arrested cracks, expected to exist in
some range of the driving $\Delta$ around the Griffith's displacement. This involves
looking for a solution of Eq. (\ref{eqn1})
with $\vec{u}(x,y)$ independent of time and with the forces on
the right hand side arising solely from the broken bonds. If
we have a system with a finite number of rows, the natural boundary condition
requires that the top and bottom rows have fixed vertical displacements of $\Delta$
and $-\Delta$ respectively. Since the entire system is linear, we can
choose to use $\Delta=1$ and rescale the breaking criterion accordingly. 

Since we are doing a numerical calculation, we must introduce 
artificial boundaries in the direction along the crack, $\hat{x}$. What we do
is cut off the range of points whose coordinates are variables and
outside of this range impose fixed asymptotic displacements. 
On the cracked side, these are just $\vec{u} = \pm \Delta \hat{y} $ for 
positive and negative $y$.
For the uncracked side, the asymptotic displacement
involves constant strain. At each of the sites that has a variable 
displacement,
we impose the two components of the (vectorial) equation of motion.
We also impose the equations of motion at the boundaries of the system.
This gives us more equations than we have variables. The coupling
of sites which contain variable displacements with the ones with
fixed displacements gives rise to inhomogeneous terms. Combining these
observations, we can write our system in the schematic form $M u - b = 0$,
with a non-square matrix $M$. The field $u$ is then determined by the
requirement that the error be minimal. In this manner, the small errors
introduced at the boundary by having to have a fixed box size are prevented 
from causing any large errors (by modes which grow exponentially away
from the edges) in the bulk of the lattice.

\begin{figure}\centerline{\includegraphics[width=3.25in]{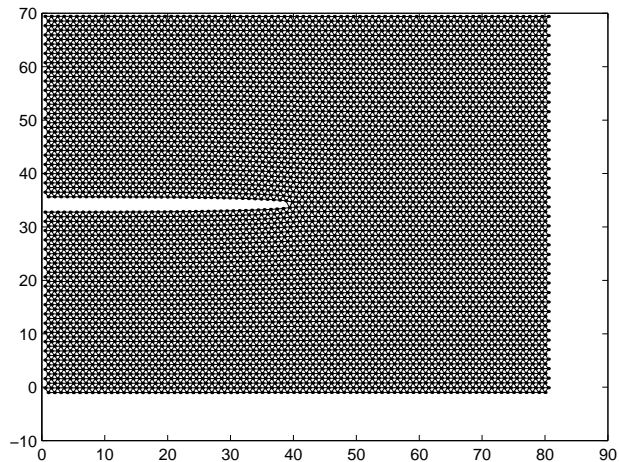}}
\caption{Finite size model for arrested crack.}
\label{figure7}
\end{figure}

\begin{figure}\centerline{\includegraphics[width=3.25in]{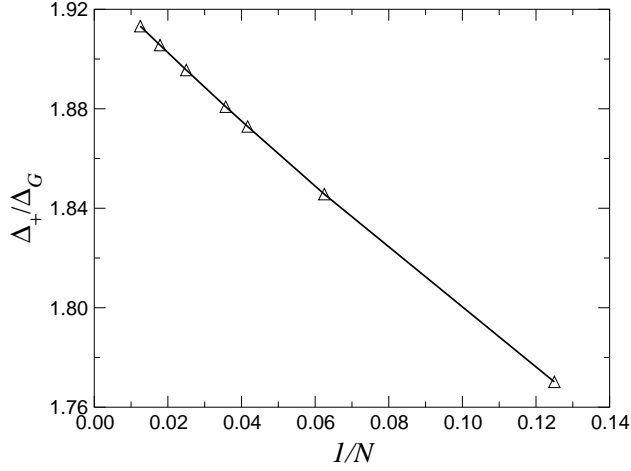}}
\caption{Mode I on triangular lattice. Upper
limiting $\Delta$  for arrested crack. Calculations were done for lattices 
with sizes $2 m\times 2 m$.}
\label{figure5}
\end{figure}

\begin{figure}\centerline{\includegraphics[width=3.25in]{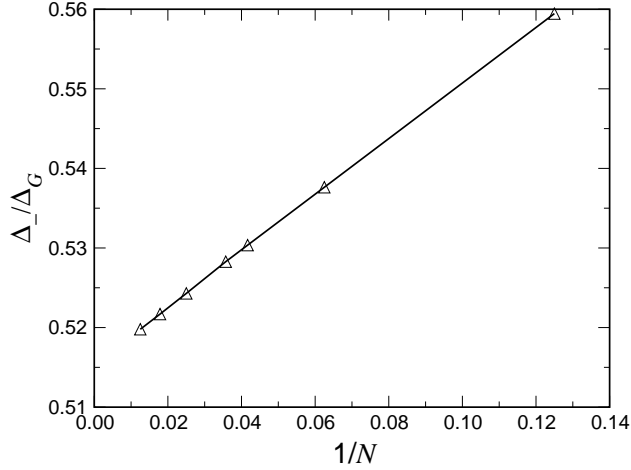}}
\caption{Mode I on triangular lattice. Lower
limiting $\Delta$  for arrested crack. Calculations were done for lattices 
with sizes $2 m  \times 2 m$.}
\label{figure6}
\end{figure}

The solution of this linear system determines the elongation of all bonds.
In Figure \ref{figure7} we show a typical lattice given by this solution. 
In fact we need to know elongation of just 
two bonds right at the tip of the crack; the one which in the case 
of the moving crack would break next ($Q_n$) and the other which 
would have been the last one broken ($Q_l$). Now, recall that we have scaled
our displacement to equal unity and the only remaining displacement scale
is the elongation at which the springs break, which we can call $\epsilon$.
The solution we have found is consistent as long as $\epsilon$ is
between the lower limit set by the $Q_n$ and the upper limit set by
$Q_l$.  Since $\Delta_{G}=\epsilon\sqrt{(2N + 1)/3}$, where $2N$ is the
number of rows in the $y$-direction (excluding the boundary rows
whose displacement is constrained), we directly
obtain the upper and lower limits of the arrested crack band
\begin{equation}
\frac{1}{\sqrt{\frac{1}{3}(2 N + 1)} Q_l}  \ \leq \ \Delta/\Delta_{G} 
\ \leq  \ \frac{1}{\sqrt{\frac{1}{3}(2 N + 1)} Q_n}
\end{equation} 
The results for these thresholds as a function
of lattice size are shown in Figs. \ref{figure5}, \ref{figure6}.
In the limit of infinite lattice, the lower limiting value 
$\Delta_-/\Delta_{G}$
approaches $0.515$, while the upper limiting value 
$\Delta_+/\Delta_{G}$ approaches $1.94$.

\subsection{Stable moving cracks}

\begin{figure}\centerline{\includegraphics[width=3.25in]{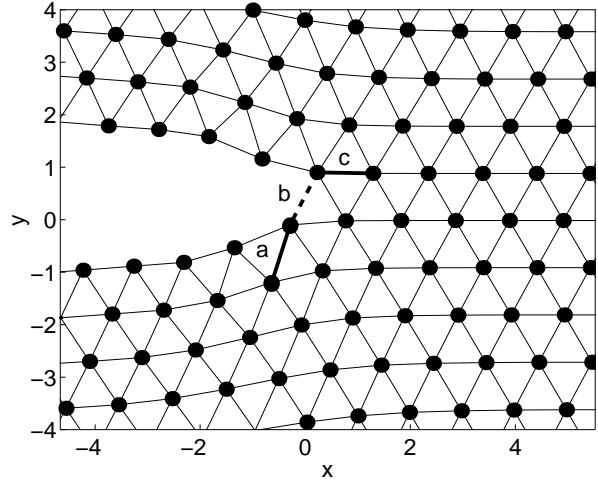}}
\caption{Snapshot of the region around tip of the crack for finite size 
model  $2N \times 2N$ for moving crack with  v=0.48, 
$\eta=1$, $n=5$, $N=24$.}
\label{figure8}
\end{figure}

We now turn to the moving crack problem. Again, we need to solve 
Eq. (\ref{eqn1}) with the
forces in the right part due  to the broken bonds. Now, the
displacement field   $\vec{u}(x,y,t)$ is of course time dependent.
Therefore, we need to introduce a time-step $\Delta t$ so as to define the
time-points at which we will obtain a numerical solution.
Given some specific speed $v$ of the crack, 
we choose to divide half of the time interval that it takes for the
crack to propagate one lattice spacing, $1/(2 v)$, into $n$ equal time 
intervals; thus  $\Delta t = 1/(2 v n)$. Now we can discretize our system,
obtaining equations  at times 
$t=0, \Delta t, ... , (n-1) \Delta t$. 
We use a symmetric discretization of first and second derivatives
$(\vec{u}(t+\Delta t )- \vec{u}(t - \Delta t ))/(2 \Delta t)$ and 
$(\vec{u}(t+\Delta t ) + \vec{u}(t - \Delta t )- 2  \vec{u}(t))
/ \Delta t ^2$ correspondingly. These equations depend on 
displacements $\vec{u}$ outside of this time interval, because of
the temporal derivatives. These can be found via use of the assumed
symmetries of the moving crack; in fact, it is easy to see that
we thereby trade in displacements outside of the modeled interval
with displacements inside this interval albeit at a different spatial
location. The boundaries are treated the same way as described for the arrested
crack; the displacements outside some range are replaced by 
asymptotic values for all of the time-points in our interval.
Including the equations at these boundary points again gives us a 
system in which the number of equations exceeds the number of variables
and again a least squared error algorithm is used to find the 
required solution.
Fig. \ref{figure8} shows a snapshot of the lattice near the tip of the crack 
at one particular time for one set of parameters.

\begin{figure}\centerline{\includegraphics[width=3.25in]{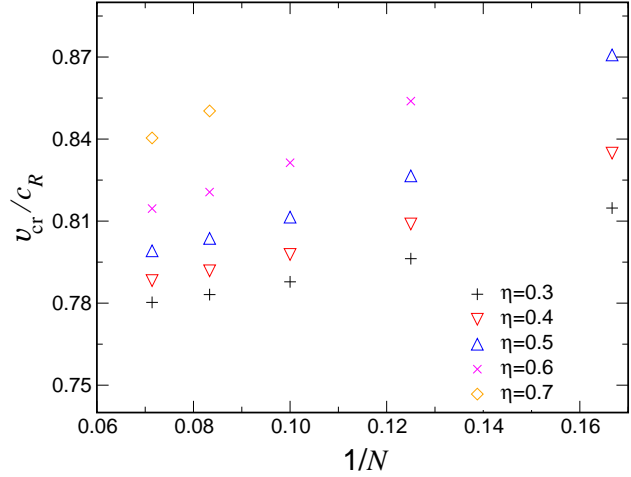}}
\caption{Critical speed due to the vertical bond braking on the layer next
to the crack. $\diamond$~ -$\eta=0.7$, $\times$ -$\eta=0.6$, $\triangle$
-$\eta=0.5$, $\bigtriangledown$ -$\eta=0.4$, $+$ -$\eta=0.3$}
\label{figure14}
\end{figure}

We used these finite lattice calculations to provide an independent check
on our analytic infinite lattice results. We were specifically interested in
checking the critical speed estimates. The difficulty is that
convergence in the parameter  $1/N$ is fairly slow; and, using large
lattices is rather time-consuming and memory-demanding. In practice,
we limited our calculations to $N$'s ranging from $6$ to $24$. 
It turns out that
for small dissipation where the next-row vertical bond vulnerability is
the thing which determines the critical velocity, we can do a credible job
of verifying that the infinite lattice results are consistent with
the finite lattice ones. For example, in 
Fig. \ref{figure14}, we show the critical 
speed as a function of lattice size for several different
values of $\eta$. These numbers, if extrapolated to infinite $N$
are clearly consistent with the results given earlier in Fig. \ref{figure12},
especially considering that having a finite time-step does introduce a
small numerical error on its own.  We can also check the critical time $\tau$
at which the bond breaking occurs. From Fig. (\ref{figure13}), we 
see that in the infinite lattice the break occurs for $\tau \sim -0.5$. For the
finite lattice with $n=5$, this mean that the bond $a$ highlighted in 
Fig. \ref{figure8} should go above breaking threshold at the first time-point 
of the modeled interval. This is exactly what occurs.

\begin{figure}\centerline{\includegraphics[width=3.25in]{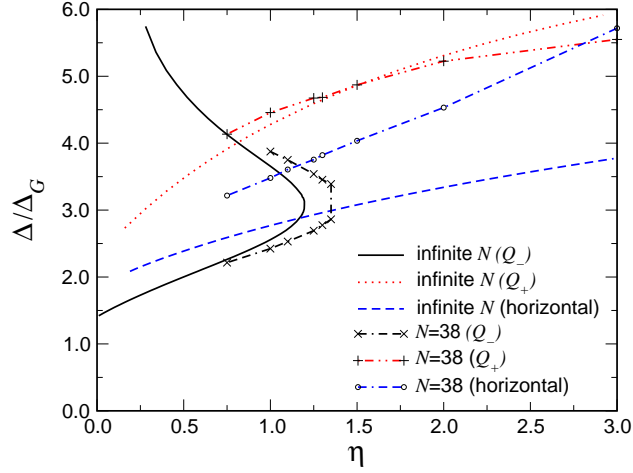}}
\caption{Comparison of the driving thresholds for critical bond extension for
$N=\infty$ and $N=38$}
\label{figure14a}
\end{figure}

\begin{figure}\centerline{\includegraphics[width=3.25in]{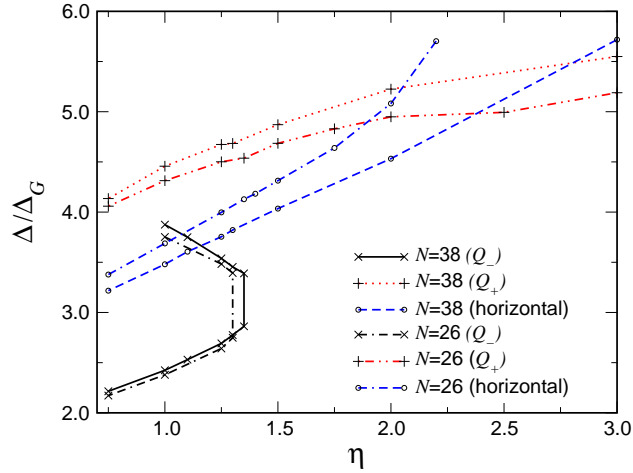}}
\caption{Comparison of the driving thresholds for critical bond extension
for $N=38$ and $N=26$.}
\label{figure14b}
\end{figure}

A different strategy is necessary for studying the onset of horizontal
bond breaking.  This is due to the fact that quite large $N$'s are required
in order to for the results to quantitatively approach the infinite $N$ limit.
This is another striking difference between the Mode I results and those
of Mode III.  For Mode III, the qualitative nature of the extraneous bond
breaking was always similar to that of the infinite width limit, and the
results were quantitatively accurate even for fairly small $N$'s.  Here,
however, the picture of which bonds are critical changes dramatically between
finite $N$ and the infinite $N$ limit.  In Fig. (\ref{figure14a}), 
we present
the ``phase diagram'' for critical bonds for $N=38$, 
plotted together with the
infinite $N$ result, and in Fig. (\ref{figure14b}), we compare the
$N=38$ results to those for $N=26$. Since the steady-state code is difficult to run at
these large values of $N$, the data in these two plots were obtained from time-dependent simulations,
with only the central vertical bonds allowed to break.  After running
to times of $t=100$ to eliminate transients, the extensions of the
dangerous bonds were checked to see if they exceeded criticality.
Note that, in contradistinction to the ``phase
diagram'' plotted in Fig. (\ref{figure12}), we now plot $\Delta/\Delta_G$
on the vertical axis, since this is the input control parameter for the
simulations.  We see that qualitative behavior of the ``$Q_-$'' bonds
is similar to the infinite-$N$ result, with these bonds being below
critical extension for both large $\eta$ and large $\Delta/\Delta_G$.
The horizontal bonds also behave qualitatively like their infinite-$N$
counterparts, but the quantitative agreement, is as we noted above, significantly worse.  In fact, for large $\eta$, the most dangerous bond is in fact
of ``$Q_+$'' type.  However, the threshold at which the ``$Q_+$'' type bond
becomes dominant is pushed to larger $\eta$ as the system size is increased,
presumably going to infinity with $N$.  Also, there is a small region of $\eta$
around 1.1 where the inconsistency is re-entrant, so that for intermediate
values of $\Delta$, no bonds of the crack surface are critical.
Thus, to see dynamics qualitatively
representative of the macroscopic system requires very large system
sizes at moderate to large $\eta$.  Presumably, this is connected to the
process zone increasing in size with $\eta$.  Again, it is worth emphasizing
that this is a feature not present in Mode III, and
leads to the conclusion that in molecular dynamics simulations, the
width of the material should be taken very large to accurately
study the micro-branching instability.

\section{Discussion}

In this work, we have discussed steady-state mode I crack propagation
in a viscoelastic lattice model. Our primary method of analysis
utilizes the Wiener-Hopf technique to write down closed form expressions
for both the $v-\Delta$ curves (for various values of the dissipation $\eta$)
and for the bond elongation field. The latter enables us to define a
limit of consistency for the solution past which some bonds not along
the crack path have elongations greater than the assumed breaking criterion.
These results are new, and correspond to non-trivial extensions of
the work of \citet{slepyan2} on the dissipationless limit and the work of
\citet{marder_gross} on small lattices.

The most interesting results, in our opinion, concern the dependence
of the critical velocity (where the aforementioned inconsistency sets in)
on the amount of dissipation. For small dissipation, this threshold
is relatively insensitive to $\eta$, as already suggested in direct
numerical simulations. This threshold, occurring at roughly .73 of
the Rayleigh speed is in some ways reminiscent of the \citet{yoffe}
branching
criterion which suggests that straight crack propagation will become
unstable once the largest stress direction shifts away from being
straight ahead. It is unfortunately hard to be more precise regarding
this correspondence since the crack dynamics in our model is fundamentally
tied to the lattice scale, not the macroscopic scale - in fact, the latter
is completely invisible in the Slepyan approach aside from providing
a driving force in the form of a stress intensity factor.

At larger $\eta$, the instability picture changes. Now, it is a horizontal
bond breaking which signals the onset of more complex crack dynamics.
Also, the threshold goes down with increasing dissipation. This was
not the case for our mode III calculations. This instability has nothing
to do with Yoffe, as it is strongly dissipation dependent and in any
case is not associated with crack branching.

Much of the recent theoretical work on mode I cracks has been motivated
by experiments which show clearly that instabilities limit the
range of steady-state crack propagation. These instabilities introduce
more complex spatio-temporal dynamics to the fracture process, causing
additional dissipation and leaving behind a roughened crack surface. It
has been tempting to associate these results with the onset of
inconsistencies in lattice models, although the details of this  
correspondence 
remains uncertain. First, most of the experimental work has been carried out
in amorphous materials, making the idea of an ordered lattice model somewhat
suspect. Next, the instabilities seen experimentally typically occur at
smaller speeds than the ones seen in lattice systems with small dissipation.
Finally, the experiments seem to show a typical frequency for micro-branching
which is not connected in any obvious way to dynamics at a lattice scale.

Notwithstanding all these issues, we remain optimistic that the study
of this class of models will lead to insight into dynamic fracture.
There are several intriguing possibilities that need to be investigated
in the future. First, we have shown that for mode I cracks, increasing the
dissipation (in the form of a Kelvin viscosity) eventually results in
a decrease of the instability threshold. If our model is really applied
at the atomic scale, it is hard to see why there should be a large linear
dissipation; on the other hand, it is well-known that lattice models
miss an essential non-linear dissipation mechanism, namely the creation
of dislocations which remain pinned to the crack. Would inclusion of
these effects also lower the threshold? On the other hand, applying the
model on a large scale (possibly for a disordered system) would naturally 
require
a large dissipation and recent numerical simulations indicate that
proper inclusion of the thermal fluctuations might also push the model into
better agreement with experiment \citep{sander1,sander2}. Finally, the exact 
nature of the state which
occurs past the instability onset has not been addressed in our work to
date, and in fact cannot be addressed by the elegant but ultimately limiting
analytic methods utilized for the steady-state problem. Instead, we plan to
study a generalized force law in which the sharp breaking criterion is
replaced by an analytic nonlinear force versus displacement \citep{arrest}.
In this formulation,
the inconsistency found here becomes a linear instability of the 
steady-state crack \citep{klnew} and one can use some of the methods developed in
the field of nonequilibrium pattern formation to unravel the dynamics 
past onset.
These studies, together with additional experimental data regarding the
differences between brittle fracture in crystalline versus non-crystalline
materials, will hopefully lead to a better understanding of dynamic fracture.

\begin{acknowledgments}
DAK acknowledges the support of the Israel Science Foundation.  The
work of HL and LP is supported in part by the NSF, grant no. DMR94-15460. DAK
and LP thank Prof. A. Chorin and the Lawrence Berkeley National Laboratory
for their hospitality during the initial phase of this work. 
\end{acknowledgments}

\newpage
\printfigures

\end{document}